\documentclass[letterpaper,preprintnumbers,english,floats,floatfix,amssymb,prd,superscriptaddress,twocolumn,nofootinbib]{revtex4-2}
\usepackage{amssymb,amsmath}
\usepackage{epsfig,hyperref}
\usepackage[svgnames]{xcolor}
\usepackage{pgf,tikz}
\usepackage{epstopdf}
\usepackage[british]{babel}
\usepackage{enumerate}
\usepackage{enumitem}
\usepackage{makecell}
\usepackage{booktabs}
\usepackage{graphicx}
\usepackage{nicefrac}
\usepackage{units}
\usepackage{mathrsfs}
\usepackage{placeins} %  

\makeatletter
\DeclareRobustCommand*{\bfseries}{%
	\not@math@alphabet\bfseries\mathbf
	\fontseries\bfdefault\selectfont
	\boldmath
}
\makeatother

\def\be{\begin{equation}}
\def\ee{\end{equation}}
\def\beq{\begin{eqnarray}}
\def\eeq{\end{eqnarray}}

\makeatletter

\newcommand{\arXiv}[2][]{\href{http://arxiv.org/abs/#2}{\texttt{arXiv:#2\@ifempty{#1}{}{ [#1]}}}}
\makeatother

\begin{document}
\title{Dimensional dependence of critical phenomena in black hole interiors}
	
\author{Li-Jie Xin}
%\email{xlijie@scut.edu.cn}
\affiliation{School of Physics and Optoelectronics, South China University of Technology, Guangzhou 510641, China}

\author{Xiangdong Zhang}
\email[Corresponding author: ]{scxdzhang@scut.edu.cn}
\affiliation{School of Physics and Optoelectronics, South China University of Technology, Guangzhou 510641, China}	
%\author{Xiangdong Zhang\textsuperscript{1}}
%\email{corresponding author:scxdzhang@scut.edu.cn}
	
\date{\today}
	
\begin{abstract}
Black hole interiors exhibit a critical transition in which a perturbed inner horizon either survives or collapses into a spacelike singularity, but whether the exponent governing this transition depends on spacetime dimension has remained an open question. In this paper, we numerically investigate the interior dynamics of $D$ dimensional Reissner–Nordstr\"om (RN) black holes perturbed by a real scalar field. Strong perturbations destroy the null inner horizon and produce a spacelike singularity, whereas weaker perturbations preserve it with an areal radius that decreases as the perturbation amplitude increases. As the initial scalar field amplitude $p$ approaches its critical value $p^{*}$, the inner horizon radius decreases continuously to zero. Near criticality, this radius obeys the power law $r_{s}\propto|p-p^{*}|^\gamma$. In four dimensions, different initial scalar field profiles and initial RN black hole charges yield nearly identical critical exponents, $\gamma\simeq1/2$, supporting the universality of the scaling law. Within the investigated range $4\leq D\leq9$, the critical exponent decreases systematically with increasing spacetime dimension. The numerical results suggest the scaling relation $r_s\propto|p-p^{*}|^{1/(D-2)}$. This result establishes a new dimension-dependent universality class for interior black hole critical phenomena, distinct from the standard Choptuik collapse scaling.
\end{abstract}
\maketitle

\section{Introduction\label{sec:introduction}}

Critical phenomena in gravitational collapse provide a framework for studying scaling behavior in black hole formation and spacetime evolution in strong gravitational fields. In 1993, Choptuik discovered critical behavior in the collapse of a massless real scalar field in four dimensional, spherically symmetric spacetime~\cite{Choptuik:1992jv}. For a family of initial data characterized by a single parameter $p$, a critical value $p^{*}$ separates scalar field dispersion from black hole formation. Near this threshold, the black hole mass in the supercritical regime obeys the power law $M_{\mathrm{BH}}\propto|p-p^{*}|^{\gamma}$, with a critical exponent $\gamma\simeq0.37$ that is independent of the family of initial scalar field profiles. This result demonstrated the universality of critical scaling for this system and motivated further studies of nonlinear gravitational collapse. Since then, critical phenomena have been investigated across a wide range of matter models and spacetime settings~\cite{Abrahams:1993wa,Hirschmann:1994du,Choptuik:1996yg,Liebling:1996dx,Ilha:1996tc,Gundlach:1996je,Hod:1996ar,Brady:1997fj,Ayal:1997ab,Koike:1999eg,Choptuik:1999gh,Markovic:1999di,Carr:1999xs,Husa:2000kr,Bizon:2001ju,Gundlach:1999cw,Olabarrieta:2001wy,Honda:2001xg,Oren:2003gp,Ventrella:2003fu,Choptuik:2003ac,Mann:2006yu,Borkowska:2011zh,Bizon:2011gg,Golod:2012yt,Nakonieczna:2012in,Maliborski:2012gx,deOliveira:2012dt,Akbarian:2014gna,Maliborski:2014rma,Bizon:2015pfa,Santos-Olivan:2015yok,Santos-Olivan:2016djn,Eilon:2015axa,Baumgarte:2019fai,Guo:2018yyt,Hatefi:2019ews,Benitez:2020szx,Guo:2020ked,Benitez:2021zjs,Jimenez-Vazquez:2022fix,Fernandez:2022hyx,Reid:2023jmr,Baumgarte:2023tdh,Baumgarte:2023saw,Marouda:2024epb,Gundlach:2024eds,Xin:2026eoy,Xin:2026aix}.

One question raised by these studies is how critical behavior depends on the spacetime dimension. Numerical evolution in higher dimensions poses significant challenges, particularly near $r=0$, where terms involving high inverse powers of the areal radius can make integration difficult. Nevertheless, higher-dimensional simulations have revealed systematic changes in critical behavior. Bland \textit{et al.}~\cite{Bland:2005kk} studied spherically symmetric scalar field collapse over the range $3.5\leq D\leq14$. They found that the critical exponent governing the horizon radius at black hole formation increases with $D$, with their numerical results well described by $\gamma(D)\simeq0.466\left(1-e^{-0.408D}\right)$. This empirical fit approaches $0.466$ at large $D$. Deppe \textit{et al.}~\cite{Deppe:2012wk} examined critical collapse in five- and six-dimensional Einstein--Gauss--Bonnet gravity. Their simulations provided evidence for a finite radius gap in five dimensions, whereas no radius gap was evident in six dimensions, where the scaling behavior changed below the Gauss--Bonnet scale.

Dimensional dependence has also been found in the geometry of critical spacetimes. Ecker \textit{et al.}~\cite{Ecker:2024haw} derived the angle $\alpha=2\operatorname{arccot}(D-1)$ between null energy condition saturation lines at the center of the critical spacetime, obtaining $\alpha\approx0.64$ in four dimensions. More recently, Mittal \textit{et al.}~\cite{Mittal:2026ryg} investigated charged scalar field collapse in $4+1$ dimensions. They found type II critical behavior characterized by self-similarity and power-law scaling of the black hole mass and charge, and provided numerical evidence for extremal black hole formation from regular initial data.

While these studies concern black hole formation and the geometry of critical solutions, the subsequent interior evolution raises questions about inner-horizon stability and singularity formation. Alongside investigations of the exterior dynamics and stability of higher-dimensional Reissner--Nordstr\"om (RN) black holes~\cite{Kodama:2003kk,Konoplya:2007jv,Konoplya:2008rq,Konoplya:2008au}, extensive work has examined the nonlinear dynamics inside RN black holes~\cite{Simpson:1973ua,Gursel:1979zza,Gursel:1979zz,Poisson:1990eh,Ori:1991zz,Gnedin:1993nau,Strominger:1993yf,Bonanno:1994ma,Levin:1996qt,Burko:1997zy,Burko:1997fc,Burko:1998jz,Hod:1998ovs,Burko:2002qr,Frolov:2005ps,Hansen:2005am,Thorlacius:2006tf,Hong:2008mw,Avelino:2009vv,Brady:1995ni,Burko:2002fv,Lin:2024beb}. In four dimensions, Brady and Smith numerically studied charged black holes perturbed by a massless scalar field~\cite{Brady:1995ni}. They found strong evidence for mass inflation and a null singularity along the Cauchy horizon, generally followed by a spacelike $r=0$ singularity deeper inside the black hole. Burko subsequently showed that the causal structure of the singularity can depend on the initial scalar field profile~\cite{Burko:2002fv}. For initial profiles characterized by a power-law exponent $\alpha$, he identified a critical value $\alpha_{\mathrm{crit}}$: the singularity contains both null and spacelike sectors for $\alpha<\alpha_{\mathrm{crit}}$, but is entirely spacelike for $\alpha>\alpha_{\mathrm{crit}}$.

Although the exterior dynamics of black holes have been extensively studied, their interiors exhibit rich nonlinear phenomena, including critical scaling associated with singularity formation. Shao \textit{et al.} numerically investigated the onset of an $r=0$ singularity inside spherically symmetric charged black holes in the Einstein--Maxwell--real scalar model~\cite{Shao:2025fki}. As the initial scalar field parameter approaches its critical value from below, the areal radius $r_s$ at the tip of the causal diamond tends to zero according to $r_s\propto|p-p^{*}|^{\gamma}$, with $\gamma\approx0.5$. The exponent is independent of both the initial scalar field profile and the size of the causal diamond, providing evidence for universality at this interior threshold.

Related radius scaling has been reported for regular Hayward black holes~\cite{Shao:2025apr}. Numerical simulations of neutral scalar field collapse in Hayward spacetime show that weak perturbations leave an inner horizon with a finite radius, whereas sufficiently strong perturbations drive its radius to zero and produce a spacelike singularity with a Schwarzschild-like interior. Near the critical value $p^{*}$ of the initial scalar field parameter, the inner horizon radius obeys $r_{-}\propto|p-p^{*}|^{\gamma}$, again with $\gamma\approx0.5$.

Interior criticality can also involve quantities other than an areal radius. Li \textit{et al.}~\cite{Li:2025mhy} identified critical scaling of the interior Kasner parameter in spontaneously scalarized black holes. Near the bifurcation of the scalarized branch from the RN solution, this parameter obeys $\beta\propto(q/q_c-1)^{-1/2}$, where $q=Q/M$ and $q_c$ is the critical charge-to-mass ratio. This scaling occurs within a family of static solutions, in contrast to the dynamical thresholds obtained by tuning initial scalar field profiles.

The interior scaling results discussed above concern four-dimensional black holes, whereas higher-dimensional interior dynamics have been investigated in the context of mass inflation. Avelino \textit{et al.} combined analytical approximations with fully nonlinear numerical simulations to study mass inflation inside higher-dimensional RN black holes with a cosmological constant~\cite{Avelino:2011ee}. They found that mass inflation persists beyond four dimensions and that the metric coefficients vary more steeply with the areal radius as the spacetime dimension increases. Their numerical results exhibited qualitative similarities to the four-dimensional case and were consistent with the homogeneous approximation.

These higher-dimensional scaling results obtained in the black hole exterior, together with the four-dimensional interior areal radius scaling described above, motivate an investigation of the dimensional dependence of interior critical behavior for the first time. In particular, does the areal radius scaling persist beyond four dimensions, and how does its critical exponent depend on the spacetime dimension? To address these questions, we numerically evolve $D$-dimensional RN black holes perturbed by a real scalar field and examine the near-critical scaling of the inner-horizon areal radius. Our four-dimensional calculations provide a benchmark against the previously reported scaling, while the higher-dimensional simulations probe how the critical exponent changes with $D$.

The paper is organized as follows. In Sec.~\ref{sec:methodology}, we present the evolution equations in double null coordinates, specify the initial data, and describe the numerical methods. Section~\ref{sec:tr} presents the results for the critical scaling of the inner-horizon areal radius. We first examine its universality in four dimensions and then determine the dimensional dependence of the critical exponent through higher-dimensional simulations. Finally, we summarize and discuss our findings in Sec.~\ref{sec:summary}. Throughout this paper, we adopt geometric units with $G=1$.

%第一节结束，第二节在这里
\section{Methodology\label{sec:methodology}}
%\subsection{polar coordinates}
We consider the Einstein-Maxwell-real scalar field model in $D$ dimensional spacetime, described by the action
\be
S = \int d^{D}x \sqrt{-g} \left(\frac{R}{16\pi G }+\mathcal{L}_{M}\right), \label{action}
\ee
where $R$ is the Ricci scalar. The matter Lagrangian density $\mathcal{L}_{M}$ for the Maxwell field and a minimally coupled massless real scalar field $\phi$ is 
\be
\mathcal{L}_{M} = -\frac{F^{2}}{16\pi}-\frac{\phi_{,a}\phi^{,a}}{8\pi},\label{mater}
\ee
where a comma denotes a partial derivative, $F^{2}=F_{ab}F^{ab}$ and $F_{ab}=\partial_{a}A_{b}-\partial_{b}A_{a}$ is the electromagnetic field strength tensor associated with the gauge potential $A_{a}$. The matter energy-momentum tensor is
\be
T_{ab} = \frac{1}{4\pi}\left(F_{ac}F^{c}_{b}-\frac{g_{ab}F^{2}}{4}+\phi_{,a}\phi_{,b}-\frac{g_{ab}\phi_{,c}\phi^{,c}}{2}\right).\label{energy-momentum}
\ee
The equations of motion obtained by varying the action in Eq.~(\ref{action}) consist of the Einstein field equations,
\be
G_{ab} = 8\pi T_{ab},\label{EOM}
\ee
and the matter field equations,
\be
\Box \phi =0,\quad d\star F = 0,
\ee
where $\Box=\nabla^a\nabla_a$ is the d'Alembertian operator, $d$ is the exterior derivative, and $\star$ represents the Hodge dual operator.

We adopt double null coordinates, in which the $D$ dimensional spherically symmetric metric takes the form
\be
ds^{2} = -2e^{2\sigma\left(u,v\right)}dudv + r^{2}\left(u,v\right)d\Omega_{D-2}^{2},\label{line_element}
\ee
where $r\left(u,v\right)$ is the areal radius and $d\Omega_{D-2}^{2}$ is the line element on the unit sphere in $(D-2)$ dimensions. The Maxwell equations then yield
\be
F_{uv}r^{D-2}e^{-2\sigma}=Q,\label{def_Q}
\ee
where $Q$ denotes the conserved electric charge. The Einstein equations and the scalar field equation reduce to
\be
r_{,uu}+\frac{2}{D-2} r \phi^{2}_{,u}-2r_{,u}\sigma_{,u}=0,\label{Einstein_uu}
\ee
\be
r_{,vv}+\frac{2}{D-2} r \phi^{2}_{,v}-2r_{,v}\sigma_{,v} =0,\label{Einstein_vv}
\ee
\be
2r r_{,uv}+\left(D-3\right)\left(e^{2\sigma}+2r_{,u}r_{,v}\right)-\frac{2e^{2\sigma}Q^{2}}{\left(D-2\right)r^{2D-6}}=0,\label{Einstein_uv}
\ee
\be
\begin{aligned}
&2r^{2}\sigma_{,uv}+2r^{2}\phi_{,u}\phi_{,v}+\frac{\left(3D-8\right)e^{2\sigma}Q^{2}}{\left(D-2\right)r^{2\left(D-3\right)}}\\
&\qquad-\frac{\left(D-3\right)\left(D-2\right)}{2}\left(e^{2\sigma}+2r_{,u}r_{,v}\right)=0,\label{Einstein_rr}
\end{aligned}
\ee
\be
\phi_{,uv}+\frac{D-2}{2}\frac{\phi_{,u}r_{,v}+\phi_{,v}r_{,u}}{r}=0.\label{KG_uv}
\ee

We define the mass function within a sphere of areal radius $r$ as
\be
M\left(u,v\right) = \frac{r^{D-3}}{2}\left(1+\frac{q^{2}}{ r^{2\left(D-3\right)}}+\frac{2 r_{,u}r_{,v}}{e^{2\sigma}}\right),\label{def_mass}
\ee
where $q$ is the charge parameter appearing in the $D$ dimensional RN black hole metric. It is related to the electric charge $Q$ by
\be
Q=\sqrt{\frac{\left(D-2\right)\left(D-3\right)}{2}}q.\label{eq:Qtoq}
\ee

For all spacetime dimensions considered, we set the initial black hole mass parameter to $m_{0}=1.0$ and the electric charge to $Q=0.95$. The outer and inner horizon radii of the unperturbed $D$ dimensional RN black hole are 
\be
r_{\pm}=\left(m_{0}\pm\sqrt{m_{0}^{2}-q^{2}}\right)^{\frac{1}{D-3}}.\label{def_horizon}
\ee

We prescribe characteristic initial data on two intersecting null rays, $u=u_{0}$ and $v=v_{0}$. The computational domain is
\be
u_{0}=0\leq u<u_{max},\qquad v_{0}=5\leq v<10.\label{eq:uvdomain}
\ee
The upper limit $u_{max}$ is adjusted for each spacetime dimension to keep regions with negative areal radius outside the computational domain. For $v<v_{0}$, the geometry is described by the RN solution. Along the initial null ray $u=u_0$, we prescribe the scalar field profile
\be
\phi(u_0,v)=
\begin{cases}
\frac{p\delta}{2}\left(x-\dfrac{\sin(2\pi x)}{2\pi}\right),
& 0\leq x<1,\\[6pt]
\frac{p\delta}{2}, & x\geq 1,
\end{cases}
\label{eq:initial_phiu0}
\ee
where $x=(v-v_0)/\delta$, while $p$ and $\delta$ control the amplitude and width of the scalar field pulse, respectively. Along the other initial null ray, $v=v_0$, we impose vanishing outgoing scalar field flux, which implies that $\phi$ is constant. We set this constant to zero:
\be
\phi\left(u,v_{0}\right) = 0.\label{eq:initial_phiv0}
\ee

Using the coordinate freedom $u\rightarrow u'(u)$ and $v\rightarrow v'(v)$, we choose
\be
r\left(u_{0},v\right) = v,\label{eq:initial_ru0}
\ee
and
\be
r\left(u,v_{0}\right) = v_{0} + u r_{,u}\left(u_{0},v_{0}\right).\label{eq:initial_rv0}
\ee
The value of $r_{,u}$ at the intersection of the initial null rays follows from the mass function in Eq.~(\ref{def_mass}):
\be
r_{,u}\left(u_{0},v_{0}\right) = \frac{e^{2\sigma\left(u_{0},v_{0}\right)}}{2}\left(\frac{2m_{0}}{r^{D-3}}-\frac{q^{2}}{r^{2\left(D-3\right)}}-1\right),\label{eq:initial_ru}
\ee
where all quantities on the right-hand side are evaluated at $(u_{0},v_{0})$. The constraint equations, Eqs.~(\ref{Einstein_uu}) and~(\ref{Einstein_vv}), give the derivatives of the initial metric function:
\be
\sigma_{,u}\left(u,v_{0}\right) = 0, \qquad \sigma_{,v}\left(u_{0},v\right) = \frac{r\phi_{,v}^{2}}{D-2}.
\ee
Finally, we use the remaining freedom to rescale $u$ to set 
\be
\sigma\left(u,v_{0}\right)=\sigma\left(u_{0},v_{0}\right)=-(\ln{2})/2.\label{eq:initial_sigma}
\ee

%The numerical algorithm employed in this paper proceeds as follows. Starting from the characteristic initial data, we evolve the coupled field equations on a double-null grid. Within each grid cell, the field variables at the fourth vertex are determined from their values at the other three vertices. Derivatives are approximated using second-order accurate finite differences. At each fixed value of $v$, the integration proceeds along $u$ to the boundary $u=u_{\max}$. We then advance to the next constant-$v$ slice and repeat the procedure until the final slice is reached.

%Near the event horizon, $|r_{,u}|$ can become large, requiring additional numerical resolution. We therefore employ adaptive mesh refinement in the $u$ direction. The grid spacing $\Delta u$ is adjusted to keep the relative change in the areal radius between adjacent grid points below $10^{-4}$. Additional grid points are inserted whenever this condition is violated. Further details of the numerical algorithm are provided in Refs.~\cite{Hansen:2005am,Avelino:2009vv,Avelino:2011ee}.

We integrate the coupled field equations using a second-order finite difference scheme, following the methods described in Refs.~\cite{Hansen:2005am,Avelino:2009vv,Avelino:2011ee}. The evolution equations can be written as
\be
Y^j_{,uv}=F^j\left(Y,Y_{,u},Y_{,v}\right),
    \qquad j=1,2,3,
    \label{eq:numerical_evolution}
\ee
where $Y^{j}=(r,\sigma,\phi)$ denotes the set of evolution variables. Starting from the characteristic initial data, we evolve these variables on a grid in the null coordinates $(u,v)$. Within each grid cell, the field values at the fourth vertex are determined from those at the other three vertices using finite difference approximations with second-order accuracy. For each fixed value of $v$, the integration proceeds along $u$ to the boundary $u=u_{\max}$. We then advance to the next constant $v$ slice and repeat the procedure until $v=v_{\max}$.

Near the event horizon, $|r_{,u}|$ can become large, requiring additional resolution to resolve the rapid variation of the fields. Since the strongest gradients occur primarily in the $u$ direction, we employ adaptive mesh refinement in this direction. The local grid spacing $\Delta u$ is adjusted to satisfy
\begin{equation}
    \frac{|\Delta r|}{r}<10^{-4},
\label{eq:refinement_criterion}
\end{equation}
where $|\Delta r|$ is the absolute change in the areal radius between adjacent grid points at fixed $v$. Additional grid points are inserted whenever this criterion is violated, allowing the grid to resolve regions with steep gradients.

For each configuration, we determine the critical amplitude $p^{*}$ by fine-tuning the initial scalar field amplitude while keeping all other parameters fixed. The fine-tuning is performed to seven significant figures, and the resulting estimate of $p^{*}$ is used in the near-critical scaling analysis.

\begin{table*}[htb]
	\centering
    \renewcommand{\arraystretch}{3.0}
    \setlength{\tabcolsep}{12pt}
	\begin{tabular}{c|c|c|c}    
		\hline
		\rule{0pt}{2.0em}Family  & Form of initial data & $Q$ & $\gamma$  \\
		\hline
			
        $(1)$ & $\phi(v) = \frac{p}{4\pi}\left[2\pi\left(v-v_{0}\right)-\delta\sin\left(\frac{2\pi\left(v-v_{0}\right)}{\delta}\right)\right]$ & $0.95$&$0.4947$  \\
		\hline
			
        $(2) $ & $ \phi(v) = p\left[\frac{1+\cos\left(\frac{2\pi\left(v-v_{c}\right)}{\delta}\right)}{2}\right]^{2}$   & $0.95$   &$0.4996$ \\
		\hline
        	
        $(3)$   &$\phi(v) = p\sin^{2}\left(\frac{\pi\left(v-v_{0}\right)}{\delta}\right)$      & $0.95$&$0.4997$ \\
        \hline
        
        $(3)$ &  $\phi(v) = p\sin^{2}\left(\frac{\pi\left(v-v_{0}\right)}{\delta}\right)$  & $0.96$&$0.4988$ \\
        \hline
        	
        $(3)$ & $\phi(v) = p\sin^{2}\left(\frac{\pi\left(v-v_{0}\right)}{\delta}\right)$  & $0.93$&$0.4988$ \\
        \hline
        	
        $(3)$  & $\phi(v) = p\sin^{2}\left(\frac{\pi\left(v-v_{0}\right)}{\delta}\right)$  & $0.90$&$0.4996$ \\
        \hline

	\end{tabular}
\caption{Initial scalar field profiles, initial RN black hole charges, and corresponding critical exponents in four dimensions. The first three rows compare different scalar field profiles with a fixed charge $Q=0.95$, while the last three rows compare different charges using the profile of Family (3). For each configuration, the scalar field amplitude $p$ is varied to determine the critical exponent $\gamma$. All critical exponents are reported to four significant figures.}
	\label{tab:initial_form}
\end{table*}

\begin{figure*}[t!]
	\centering
	\begin{tabular}{cc}
		\includegraphics[width=1.0\textwidth]{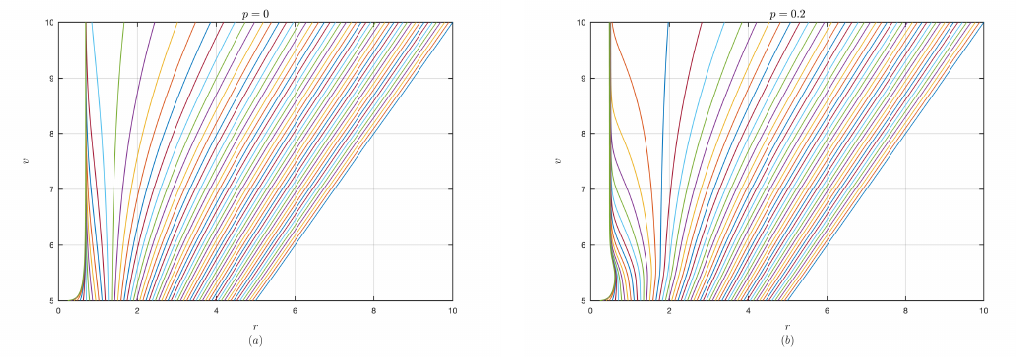}
	\end{tabular}
	\caption{Evolution of the areal radius $r$ as a function of the null coordinate $v$ in a four dimensional RN black hole spacetime: (a) the unperturbed case ($p=0$); (b) the perturbed case ($p=0.2$ and $\delta=1$). Each curve corresponds to a null ray at fixed $u$.}
	\label{fig:r_vs_v}
\end{figure*}

\begin{figure*}[t!]
	\centering
	\begin{tabular}{cc}
		\includegraphics[width=0.9\textwidth]{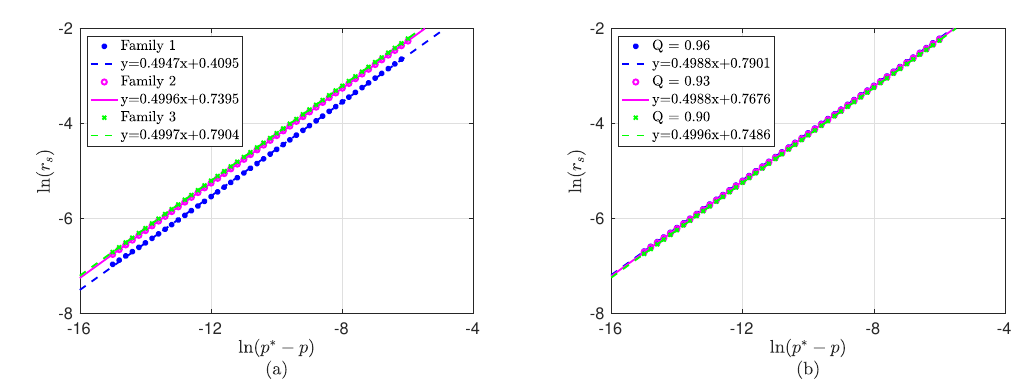}
	\end{tabular}
	\caption{Universality of the power-law scaling of the inner horizon areal radius with the deviation of the initial scalar field amplitude from its critical value. The critical exponent remains close to $1/2$ across the initial scalar field profiles and RN black hole charges considered. (a) Scaling relations for different initial scalar field profiles. The blue dashed, magenta solid, and green dashed curves correspond to Families (1), (2), and (3), respectively. (b) Scaling relations for different initial RN black hole charges. The blue dashed, magenta solid, and green dashed curves correspond to $Q=0.96$, $Q=0.93$, and $Q=0.90$, respectively. The profiles and charges used are listed in Tab.~\ref{tab:initial_form}.}
	\label{fig:initial}
\end{figure*}

\begin{figure*}[t!]
	\centering
	\begin{tabular}{cc}
		\includegraphics[width=0.99\textwidth]{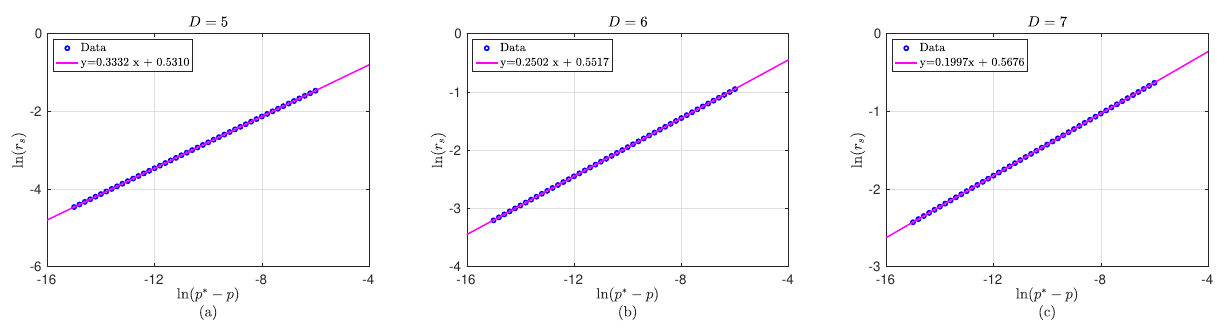}
	\end{tabular}
	\caption{Power-law scaling of the inner horizon areal radius of RN black holes with the deviation of the amplitude from its critical value. Panels (a), (b), and (c) show the scaling relations in five, six, and seven spacetime dimensions, respectively.}
	\label{fig:D}
\end{figure*}

\begin{table}[htb]
	\centering
	\begin{tabular}{c|c}    
		\hline
		\rule{0pt}{1.5em}spacetime dimension ($D$)  &  Critical exponent ($\gamma$) \\
		\hline
		\rule{0pt}{1.6em}	$ 4$   &$0.4947\pm 0.0015$  \\
		\hline
		\rule{0pt}{1.6em}	$5$     &$0.3332\pm 0.0001 $  \\
		\hline
		\rule{0pt}{1.6em}	$6$ &$0.2502\pm 0.0002  $  \\
		\hline
        \rule{0pt}{1.6em}	$7$    &$0.1997 \pm 0.0001 $  \\
        \hline
        \rule{0pt}{1.6em}	$8$   &$0.1662 \pm 0.0002 $  \\
        \hline
        \rule{0pt}{1.6em}	$9$  &$0.1417 \pm 0.0003 $  \\
        \hline
       
	\end{tabular}
\caption{Critical exponent $\gamma$ for RN black holes in different spacetime dimensions. Each exponent characterizes the power-law scaling of the inner horizon areal radius with the deviation of the initial scalar field amplitude from its critical value. All values are reported to four significant figures.}
	\label{tab:case1}
\end{table}

\begin{figure}[t!]
	\centering
	\begin{tabular}{cc}
		\includegraphics[width=\columnwidth]{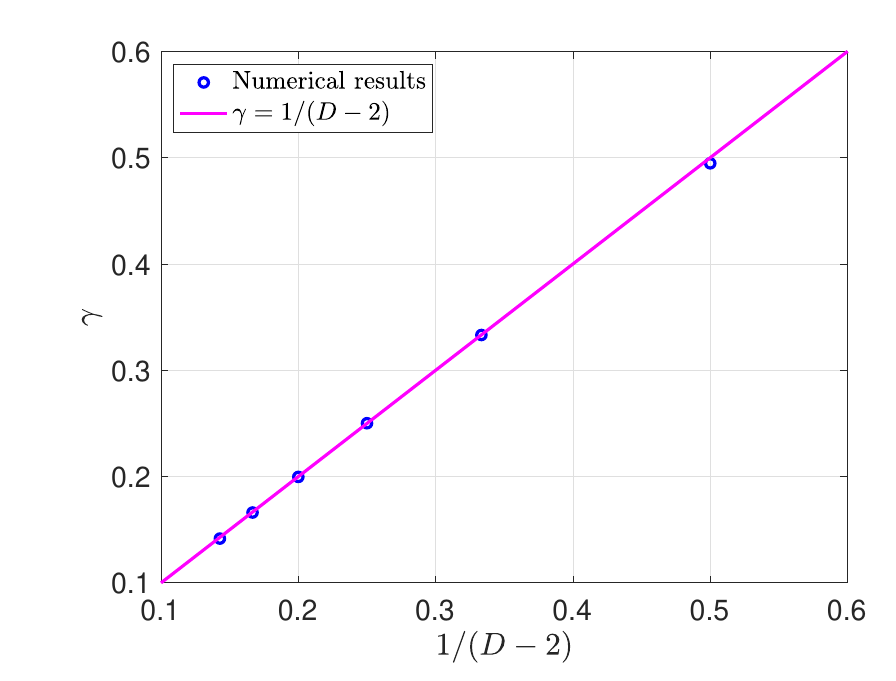}
	\end{tabular}
	\caption{Critical exponent $\gamma$ as a function of the spacetime dimension $D$ for the power-law scaling of the RN black hole inner horizon areal radius.}
	\label{fig:dimen}
\end{figure}

%\begin{figure*}[t!]
%	\centering
%	\begin{tabular}{cc}
%		\includegraphics[width=1.0\textwidth]{fig4_dyna.pdf}
%	\end{tabular}
%	\caption{{\color{blue}Contribution of each term in the equation for the scalar field with potential~(\ref{potential}) at $x = 0.0125$. (a) Contributions of the terms in Eq.~(\ref{con_f}). (b) Contributions of the terms in Eq.~(\ref{con_tx}). (c) Approximate results in Eq.~(\ref{con_tx}), where $a/2\approx r_{,x} \approx r/x$.}}
%	\label{fig:terms}
%\end{figure*}

%\begin{figure*}[t!]
%	\centering
%	\begin{tabular}{cc}
%		\includegraphics[width=1.0\textwidth]{fig5_poten.pdf}
%	\end{tabular}
%	\caption{{\color{blue}Contribution of each term in Eq.~(\ref{con_f}) for different forms of the scalar field potential with $\ell=8$ at $x=0.0125$. (a) $V\left(\phi\right)=\mu^{2}\phi^{2}/2$, $\mu=1.0$. (b) $V\left(\phi\right)=\lambda\phi^{4}/4$, $\lambda=1.0$. (c) $V\left(\phi\right)=\lambda\phi^{4}/4$, $\lambda=-1.0$.} }
%	\label{fig:poten}
%\end{figure*}

%\begin{figure}[t!]
%	\centering
%	\begin{tabular}{cc}
%		\includegraphics[width=0.45\textwidth]{fig2_prltype1.pdf}
%	\end{tabular}
%	\caption{Relationship between the black hole mass and the initial amplitude in the supercritical regime for a scalar field with potential $ V \left(\phi \right) = \mu^{2}\phi^{2}/2 $ and mass parameter $\mu = 8$. The blue stars and magenta circles correspond to the cases $k=0$ and $k=1$, respectively.}
%	\label{fig:prltype}
%\end{figure}

\section{critical phenomena in black hole interiors \label{sec:tr}}
%\subsection{Type II critical collapse$}
We begin with the unperturbed four dimensional RN black hole, corresponding to $p=0$. Fig.~\ref{fig:r_vs_v}(a) shows the evolution of the areal radius $r$ with the null coordinate $v$ along null rays at different fixed values of $u$. As $v$ increases, the areal radius $r$ inside the black hole approaches a finite value of approximately $0.6877$, in good agreement with the theoretical inner horizon radius given by Eq.~(\ref{def_horizon}). This agreement provides a consistency check of the numerical implementation and its ability to reproduce the inner horizon structure of the stationary RN solution. The unperturbed evolution also serves as a reference for identifying the changes induced by scalar field perturbations.

Fig.~\ref{fig:r_vs_v}(b) shows the corresponding evolution for a scalar field perturbation with $p=0.2$ and $\delta=1$. In this case, the areal radius $r$ inside the black hole approaches a smaller finite value of approximately $0.4751$. Thus, the perturbation preserves an inner horizon with a finite areal radius while substantially reducing its minimum areal radius inside the black hole relative to the unperturbed case. The comparison between the two panels illustrates how the scalar field modifies the interior geometry, with the limiting inner horizon radius providing a useful measure of this response. In the following, we denote the characteristic areal radius of the perturbed inner horizon by $r_{s}$.

As the perturbation amplitude increases, $r_{s}$ decreases continuously toward zero. For perturbation amplitudes above a critical value, the null inner horizon at finite radius is destroyed, and a spacelike singularity forms at $r=0$. These outcomes define a threshold separating interiors with a surviving null inner horizon from those in which it is replaced by a spacelike singularity. To characterize the approach to this threshold, we take the scalar field amplitude $p$ as the control parameter and denote its critical value by $p^{*}$. We approach $p^{*}$ from the side on which the inner horizon retains a nonzero radius. By fine-tuning the amplitude, we obtain $46$ data points uniformly spaced in $\ln|p-p^{*}|$ over the interval $[-15,-6]$. A linear fit in logarithmic coordinates yields
\be
r_{s}\propto|p-p^{*}|^{\gamma},
\label{eq:powerlaw}
\ee
with $\gamma\approx0.4947$. This independently reproduces the four-dimensional scaling behavior reported in Ref.~\cite{Shao:2025fki} and provides a consistency check of our numerical method. Building on this benchmark, we extend the analysis to higher dimensions to investigate the dimensional dependence of the critical exponent, consistent with the behavior reported in Ref.~\cite{Shao:2025fki}. We obtain $\gamma\approx0.4947$, reproduces the benchmark reported in Refs.~\cite{Shao:2025fki,Shao:2025apr}. The power-law behavior provides a quantitative description of how the inner horizon radius vanishes as the initial data approach the critical threshold.

To test the universality of this scaling relation, we repeat the four dimensional simulations with different initial scalar field profiles. For each family, we hold all profile parameters except the amplitude fixed and fine-tune the amplitude toward its corresponding critical value. The resulting scaling relations are shown in Fig.~\ref{fig:initial}(a), and the initial scalar field profiles are listed in Tab.~\ref{tab:initial_form}. Despite the differences among these profiles, the fitted critical exponents are nearly identical, all remaining close to $1/2$. The critical amplitude and the intercept of the linear fit in logarithmic coordinates vary with the initial data family, whereas the critical exponent remains approximately unchanged. These findings support the universality of the critical exponent across the different initial scalar field profiles considered.

We further examine the dependence of the critical scaling on the initial charge of the RN black hole by performing additional simulations with different initial charges. As shown in Fig.~\ref{fig:initial}(b), the power-law scaling persists, with the fitted critical exponents consistently remaining close to $1/2$. Together, these results provide numerical evidence that the critical exponent is approximately independent of both the initial scalar field profile and the initial black hole charge over the range explored.

We next extend the analysis to higher dimensional RN black holes. In the remainder of this paper, we adopt the scalar field profile of Family (1). We investigate whether the areal radius of the null inner horizon continues to exhibit power-law scaling and how the corresponding critical exponent depends on the spacetime dimension $D$. This allows us to explore the role of spacetime dimensionality in the dynamical evolution of the null inner horizon near the threshold of its disappearance.

We first consider the five dimensional case. As the amplitude approaches its critical value, the inner horizon radius again follows a power law, as shown in Fig.~\ref{fig:D}(a). A linear fit in logarithmic coordinates yields $\gamma\approx0.3332$, close to $1/3$. Compared with the four dimensional result, the scaling form is preserved, while the critical exponent decreases. Similar behavior is found in six and seven dimensions, as illustrated in Figs.~\ref{fig:D}(b) and~\ref{fig:D}(c), respectively. In each case, the approximately linear relation in logarithmic coordinates indicates that the approach to a vanishing inner horizon radius remains well described by Eq.~(\ref{eq:powerlaw}).

We extend the simulations to $D=9$. As the spacetime dimension increases, numerically simulating the dynamical evolution near $r=0$ becomes increasingly challenging, limiting our present analysis to $4\leq D\leq9$. Within this range, power-law scaling is observed in every dimension considered, and the critical exponent decreases systematically with increasing $D$. The persistence of the scaling form across these dimensions provides numerical evidence that this critical behavior extends to higher dimensional charged black hole interiors.

To quantify the dimensional dependence, we plot the fitted critical exponents as a function of $D$ in Fig.~\ref{fig:dimen}. The numerical results are well described by
\be
\gamma\approx\frac{1}{D-2}.
\label{gamma}
\ee
Equivalently, the product $(D-2)\gamma$ remains approximately unity over the dimensional range investigated. The relation $\gamma\approx\frac{1}{D-2}$ reduces to the previously reported values $\gamma\approx1/2$ at four dimensions, and reproduces all six measured exponents for $4 \leq D \leq 9$ to within about $1\%$, without any adjustable parameter. Motivated by this agreement, we write the dimensional scaling as
\be
r_s\propto|p-p^{*}|^{\frac{1}{D-2}}.
\label{rspower}
\ee
which we regard as an empirical relation for the configurations studied here. Rearranging Eq. \ref{rspower} gives the equivalent statement
\be
|p-p^{*}|\propto r_s^{D-2}.
\label{area}
\ee
that is, the deviation of the initial scalar field amplitude from its critical value is proportional to the cross-sectional area of the surviving inner horizon. Written in this form the relation admits a suggestive geometric reading: the exponent $D-2$ is precisely the dimension of the spherical cross sections of the inner horizon, so that criticality is approached linearly in the horizon area. Whether this coincidence reflects a deeper dynamical mechanism is not established by the present data; we therefore present Eq. \ref{rspower} as an empirical scaling relation whose origin remains to be identified analytically.

%As the amplitude approaches the critical threshold, this area decreases continuously to zero, providing a geometric characterization of the approach to criticality. 

%\begin{figure*}[t!]
%	\centering
%	\begin{tabular}{cc}
%		\includegraphics[width=0.85\textwidth]{fig3_unvplot.pdf}
%	\end{tabular}
%	\caption{Critical behavior of a massive scalar field with potential $V\left(\phi\right) = \mu^{2}\phi^{2}/2$ for $\mu=1$. (a) Evolution of the scalar field $\varphi$ at the origin $(x = 0)$ as a function of $-\ln{(T^{*}-T)}$, where $T^{*}$ denotes the time of naked singularity formation.  (b) Power-law scaling of the black hole mass in the supercritical regime. The initial profile parameters in Equation~(\ref{initial_psi2}) are $x_{0}=1$, and $\sigma=0.2$. The blue solid line and the magenta dashed line correspond to the cases $k=0$ and $k=1$, respectively.}
%	\label{fig:unvslop}
%\end{figure*}

%\begin{figure}[t!]
%	\centering
%	\begin{tabular}{cc}
%		\includegraphics[width=0.45\textwidth]{fig4_unvtype1.pdf}
%	\end{tabular}
%	\caption{Relationship between the black hole mass and the initial amplitude in the supercritical regime for a scalar field with potential $ V \left(\phi \right) = \mu^{2}\phi^{2}/2 $ and mass parameter $\mu = 8$. The blue stars and magenta circles correspond to the cases $k=0$ and $k=1$, respectively.}
%	\label{fig:unvtype}
%\end{figure}

%\section{Result II: double null coordinates\label{sec:uv}}

\section{Summary\label{sec:summary}}
Investigating the dynamics of black hole interiors provides insight into the stability of inner horizons and the formation of spacetime singularities. In this paper, we have numerically investigated the nonlinear interior dynamics of $D$ dimensional RN black holes perturbed by a massless real scalar field. By evolving the coupled Einstein-Maxwell-scalar field equations in double null coordinates, we have examined how the perturbation amplitude and spacetime dimension affect the inner horizon structure. For weak perturbations, the null inner horizon survives with a finite areal radius that decreases as the amplitude increases. Sufficiently strong perturbations, on the other hand, destroy the null inner horizon, producing a spacelike singularity at $r=0$.

By tuning the initial scalar field amplitude toward its critical value, we find that the characteristic inner horizon areal radius $r_{s}$ approaches zero continuously and obeys the power law $r_{s}\propto|p-p^{*}|^{\gamma}$. In four dimensions, we obtain a critical exponent $\gamma\approx0.4947$, close to $1/2$. Although the critical amplitude and the intercept of the linear fit in logarithmic coordinates depend on the family of initial data, different initial scalar field profiles and initial RN black hole charges yield nearly identical critical exponents. These results provide numerical evidence for the universality of the critical exponent across the families of initial data and charge values examined.

The power-law scaling persists for the configurations studied at integer spacetime dimensions $4\leq D\leq9$, with the fitted critical exponent decreasing systematically as $D$ increases. Within this range, the numerical results suggest the simple candidate relation
\be
\gamma(D)\simeq\frac{1}{D-2}.
\ee
This expression provides an approximate description of the available data, but its validity beyond the investigated range remains to be tested.

The simplicity of this inverse dependence is noteworthy in comparison with the empirical exponential fit $0.466(1-e^{-0.408D})$ reported by Bland \textit{et al.}~\cite{Bland:2005kk} for the horizon-radius exponent at black hole formation. Although the two exponents describe different physical thresholds, the particularly simple form suggested by our interior simulations may reflect an underlying geometric or dynamical mechanism. The appearance of $D-2$, the dimension of the spherical horizon cross sections, provides a possible clue to such a mechanism. However, the present numerical evidence is restricted to $4\leq D\leq9$. The broader validity of this relation and its underlying dynamical origin remain to be investigated.

\section*{Acknowledgments}\small
This work is supported by the National Natural Science Foundation of China (NSFC) with Grant No.12275087.

\FloatBarrier	
%%%%%%%%%%%%%%%%%%%%%%%%%%%%%%%%%%%%%%%%%%%%%%%%%%%%%%%%%%%%%%%%

\end{document}